\documentclass[%
 reprint,
amsmath,
amssymb,
pra,
floatfix,
balancelastpage,
]{revtex4-1}
\usepackage{color}
\usepackage{epstopdf}
\usepackage{graphicx}
\usepackage{dcolumn}
\usepackage{bm}
\usepackage{amssymb}   
\usepackage{amsmath}
\usepackage{braket}
\usepackage{csquotes}
\usepackage{hyperref}

\usepackage{physics}
\usepackage{lipsum}

\begin{document}


\preprint{APS/123-QED}

\title{Surface polaritonic solitons and breathers in a planar plasmonic waveguide structure via electromagnetically induced transparency}

\author{Koijam Monika Devi}
 \email{koijam@iitg.ac.in}
 \author{Gagan Kumar}%
 
\author{Amarendra K. Sarma}%
 \email{aksarma@iitg.ac.in}
\affiliation{
 Department of Physics, Indian Institute of Technology  Guwahati, Guwahati-781039, Assam, India}


%

\date{\today}

\begin{abstract}
We propose a scheme for the coupler-free excitation of surface polaritonic solitons and breathers in a planar plasmonic waveguide structure comprising of a transparent layer, a metal layer and a layer of three-level lambda-type atomic medium. In the proposed system, an enhanced Kerr nonlinearity is achieved via electromagnetically induced transparency (EIT) in the bottom atomic medium, which can be controlled through proper modulation of the frequency detunings and Rabi frequencies of the driving laser fields. This nonlinearity balances the dispersion in the system, thus providing the necessary condition for the excitation of polaritonic solitons in the proposed system. As a result, the system yields laterally self-trapped bright and dark surface polaritonic solitons which are tightly guided at the interface of the metal and the EIT medium. Furthermore, we investigate the excitation of surface polaritonic Akhmediev breathers at the metal-EIT medium interface in the proposed system. A stable propagation of the surface polaritonic Akhmediev breathers is  achieved with proper choice of parameters in the proposed planar plasmonic waveguide structure. This experimentally feasible scheme could be significant in the development of highly compact nano-photonic devices in the optical regime.

\begin{description}

\item[PACS numbers]
\end{description}
\end{abstract}

\maketitle
\section{\label{intro}Introduction}


Over the past decades, the electromagnetically induced transparency (EIT) phenomenon has garnered a lot of attention due to its potential applications in the efficient control of optical properties in atomic media \cite{fleischhauer2005electromagnetically,marangos1998electromagnetically,lukin2001controlling,phillips2001storage,li1996enhancement,braje2004frequency}. EIT is a quantum phenomenon that usually occurs as a result of destructive interference between two transition pathways of an atomic medium, resulting in the elimination of the absorption in the medium \cite{fleischhauer2005electromagnetically,marangos1998electromagnetically}. It should be noted that the optical response of the medium is greatly modified within the EIT transparency window \cite{schmidt1996giant}. Numerous interesting studies have been performed, to utilize the key aspects of the EIT phenomenon such as the tunable transparency window, steep dispersion, enhanced nonlinearity, etc. for the realization of various applications such as slowing of light \cite{lukin2001controlling,phillips2001storage} and nonlinear processes \cite{braje2004frequency,li1996enhancement}, etc. A plethora of theoretical and experimental studies have also been performed to study the optomechanical \cite{wang2014optomechanical,safavi2011electromagnetically} and metamaterial analogues \cite{zhang2008plasmon,papasimakis2008metamaterial,devi2017plasmon,devi2018plasmon,mahat2017plasmonically} of the EIT phenomenon. Recently, the EIT phenomenon has also been associated with the generation of surface plasmon (SP) resonances in planar waveguide structures \cite{du2012quantum,shen2014electromagnetically}. SPs are electromagnetic surface modes enabling effective localization of light over subwavelength dimensions \cite{maier2007plasmonics}, and have been studied extensively over the years, owing to its various applications in subwavelength control of light \cite{shalaev2006nanophotonics,gramotnev2010plasmonics,schuller2010plasmonics,poudel2018active}, bio-sensing \cite{wu2010highly}, solar cells \cite{zhang2012surface,catchpole2008plasmonic,ferry2010light}, etc. The unification of the EIT phenomenon and SPs excitation has paved the way for several studies in the recent years. Theories have been proposed for the generation of SP resonances in a system of a prism-coupler with a lambda-type EIT medium \cite{du2012quantum} as well as a four-level tripod EIT system \cite{shen2014electromagnetically}. The propagation of ultraslow SPs have also been examined in a recent study, in a grating coupled planar waveguide structure via the EIT phenomenon \cite{ziemkiewicz2018ultraslow}. 

In 2015, a novel study performed by Du \textit{et.al} revealed the possibility of exciting SP resonances without the use of any coupler at the interface of a metal and an EIT medium \cite{du2015coupler}. Other studies have also been performed by Bai \textit{et. al}, in which they reported the excitation of plasmon-solitons \cite{bai2015giant} and dromions \cite{bai2016plasmon} in metamaterials (MMs) by utilizing the analog of EIT phenomenon in MMs. Recently, a coupler free  scheme for SPs excitation at the interface of a negative index metamaterial (NIMM) and a four-level EIT medium has also been studied, along with the behavior of the SPs in the non-linear regime \cite{asgarnezhad2017coupler,asgarnezhad2018excitation}. In the non-linear regime, the dynamics of the excited SPs is governed by the so-called non-linear Schrodinger equation (NLSE), which can assume different exact solutions (e.g. solitons, breathers and rogue waves) \cite{boyd2003nonlinear,drazin1989solitons}. The SPs in the non-linear regime, inherit the properties of these nonlinear waves such as stable propagation, shape preservation, modulation instability \cite{feigenbaum2007plasmon,marini2011stable,kumar2017spatial,devi2018surface}, etc. Hence, there is a significant potential of such nonlinear SP waves as it can be employed as a means to reduce the propagation loss inherent in SPs. Additionally, the coupler-free scheme can become a promising technique for the development of highly compact nano-optical devices for various applications in the optical regime by overcoming the limitation faced by the SP coupling techniques. However, only few studies on coupler-free excitation of the nonlinear SP waves has been reported so far \cite{asgarnezhad2017coupler,asgarnezhad2018excitation,dong2018matching}. Therefore, more studies are required to have a better understanding of the non-linear dynamics of the surface polaritonic solitons and breathers in such coupler free schemes. 

In this article, we propose a scheme for the coupler-free excitation of surface polaritonic solitons and breathers in a planar plasmonic waveguide structure comprising of a transparent layer, a metal layer and a three level lambda-type EIT medium. Although, similar studies have been performed earlier \cite{asgarnezhad2017coupler,asgarnezhad2018excitation}, the nonlinear behavior of the SPs using a coupler-free scheme with a three-level EIT medium has not been investigated. Further, we have used a metal layer in our study, unlike the studies reported in ref. \cite{asgarnezhad2017coupler,asgarnezhad2018excitation} where a NIMM layer is used. To the best of our knowledge, the excitation and propagation of the surface polaritonic solitons and breathers remained to be explored in our proposed structure. Here, we explore the nonlinearity aspect associated with the EIT phenomenon to investigate the generation of surface polaritonic solitons and breathers in the proposed structure. It is observed that within the EIT transparency window, a giant Kerr nonlinearity is achieved which can be efficiently controlled through proper modulation of the parameters of the incident fields. A balance between the group velocity dispersion (GVD) and the Kerr nonlinearity in the system provides the necessary condition for the excitation of nonlinear SP waves. It is observed that the system yields laterally self-trapped surface polaritonic solitons which is tightly guided at the interface of the metal and the EIT medium. Further, we explore the possibility of generating surface polaritonic breathers in the proposed system. We show that the surface polaritonic solitons and breathers can have an undistorted and a stable propagation at the interface of the metal-EIT medium. The paper is organized as follows: In section 2, the theoretical model of the coupler free scheme based on EIT is explained. The coupler free excitation of SPs in the proposed structure is discussed in section 3. Section 4 explains the excitation and propagation of the surface polaritonic solitons in the nonlinear regime. In section 5, the excitation and propagation of the surface polaritonic Akmediev breather is discussed followed by a brief conclusion in section 6.

\section{Theoretical model}
\begin{figure}[htbp]
	\centering
	\includegraphics[width=0.85\linewidth]{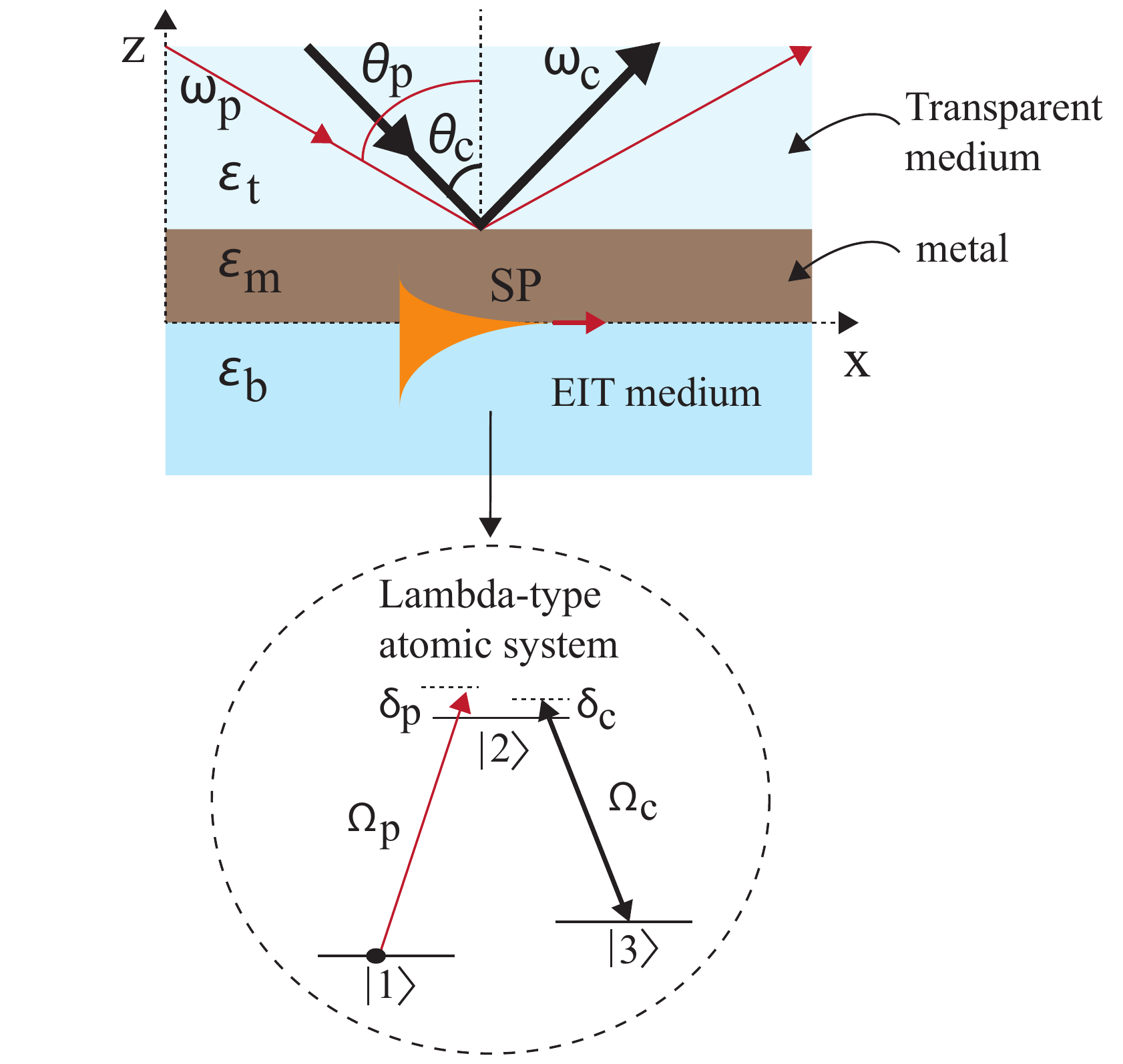}
	\caption{Schematic illustration for the coupler free excitation of surface polaritonic solitons and breathers in a planar plasmonic waveguide structure based on EIT. A three-level lambda-type atomic medium is considered as the EIT medium in the system. The red arrow indicates the probe field while the black arrow denotes the coupling field.} 
\end{figure}

The schematic diagram of the proposed planar plasmonic waveguide structure based on EIT is illustrated in Fig. 1. The structure comprises of three layers namely, a top transparent layer, a middle metal layer and a bottom layer of EIT medium. The top layer is a transparent medium (vacuum or a lossless dielectric) having a refractive index, $n_{t}= \sqrt{\epsilon_{t} \mu_{t}/\epsilon_{0} \mu_{0}}\approx 1$, where $\epsilon_{t}$ and $\mu_{t}$ are the electrical permittivity and magnetic permeability of the medium, respectively. The middle layer is considered to be a metal having a permittivity, $\epsilon_{m}$ while the bottom layer is a three-level lambda type EIT medium with electric permittivity, $\epsilon_{b}$. The levels $\ket{1}$, $\ket{2} $ and $ \ket{3}$  denotes the energy levels of the lambda-type atomic system in which the transition between the levels $\ket{1}$ and  $\ket{2}$ is driven by a weak probe field of angular frequency, $\omega_{p}$ and the transition between the levels $\ket{2}$  and  $\ket{3}$ is driven by a strong coupling field of angular frequency, $\omega_{c}$. $ \Omega_{p}= \frac{\vec{E}_{p}(\vec{r}, t).\vec{\mu}_{21}}{\hbar}$ and $\Omega_{c}=  \frac{\vec{E}_{c}(\vec{r}, t).\vec{\mu}_{23}}{\hbar}$  denotes the Rabi frequencies associated with the probe and coupling fields with frequency detunings $\delta_{p}=\omega_{p}-\omega_{21}$ and $\delta_{c}=\omega_{c}-\omega_{23}$, respectively. The dynamics of the three-level lambda-type atomic medium is given by the Maxwell Bloch equations \cite{wang2001enhanced,van2015eit}. Using the dipole and the rotating wave approximations, the equation of motion of the medium can be described by the density matrix equations as follows: 
\begin{subequations}\label{eq:1}
	\begin{align} 
	\dot{\rho_{11}} &=\gamma_{31}(\rho_{33}-\rho_{11}) +\gamma_{21}\rho_{22}- \frac{i}{2}\Omega_{p} \rho_{21}+\frac{i}{2}\Omega_{p} \rho_{12},
	\\
	\dot{\rho_{22}} &=-(\gamma_{31}+\gamma_{21})
	\rho_{22}- \frac{i}{2}\Omega_{p} \rho_{12}+\frac{i}{2}\Omega_{p} \rho_{21}-\frac{i}{2}\Omega_{c} \rho_{32}+\frac{i}{2}\Omega_{c} \rho_{23},
	\\
	\dot{\rho_{33}} &=\gamma_{31}(\rho_{11}-\rho_{33}) +\gamma_{23}\rho_{22}- \frac{i}{2}\Omega_{c} \rho_{32}-\frac{i}{2}\Omega_{c} \rho_{23},
	\\
	\dot{\rho_{21}} &=-(\gamma-i\delta_{p})\rho_{21}+\frac{i}{2}\Omega_{p} (\rho_{22}-\rho_{11})-\frac{i}{2}\Omega_{c} \rho_{31},	
	\\
	\dot{\rho_{23}} &=-(\gamma-i\delta_{c})\rho_{23}+\frac{i}{2}\Omega_{c} (\rho_{22}-\rho_{33})-\frac{i}{2}\Omega_{p} \rho_{13},	
	\\
	\dot{\rho_{31}} &=-[\gamma_{31}-i(\delta_{p}-\delta_{c})]\rho_{31}+\frac{i}{2}\Omega_{p}\rho_{32}+\frac{i}{2}\Omega_{c} \rho_{21}.		
	\end{align}
\end{subequations} 
Under the slowly varying approximation, the evolution of the weak probe field in the system which is given by the Maxwell Equation, $\nabla^{2} \vec{E}- \Big( \frac{1}{c^{2}}\Big)\frac{\partial^{2}\vec{E}}{\partial t^{2}} = \Big(\frac{1}{\epsilon_{0}c^{2}}\Big)\frac{\partial^{2}\vec{P}}{\partial t^{2}} $ as 
\begin{equation} \label{eq:2}
i \Bigg( \frac{\partial}{\partial x} +  \frac{1}{n_{eff}c^{2}}\frac{\partial}{\partial t} \Bigg) \Omega_{p} (\vec{r}, t)+\kappa \rho_{21},
\end{equation}
where, $\kappa=2 \pi N \omega_{p}{|\vec{\mu}_{21}|}^{2}/\hbar c$ is the coupling coefficient. $N$ is the atomic density of the atomic medium, $\vec{\mu}_{ij}$ is the dipole moment of the transition $\ket{j} \rightarrow \ket{i}$   (for $i, j = 1, 2, 3$), $\epsilon_{0}$ is the absolute electric permittivity. $n_{eff}= c k_{p}/ \omega_{p}$ is the effective refractive index of the system. 
Here, the dielectric constant of the medium can be obtained with effective-medium theory, as
\begin{equation}\label{eq:3}
\epsilon_{b}=1+\frac{\chi_{p}}{1-\frac{\chi_{p}}{3}},
\end{equation} 
where, $\chi_{p}$ is the effective susceptibility of the medium which is expanded to the third order, \cite{bass1995handbook} as follows: 
\begin{equation} \label{eq:4}
\chi_{p}= {\chi^{(1)}_{p}}+ \frac{3}{4}{E^{2}_{p}}{\chi^{(3)}_{p}}.
\end{equation}
\begin{figure}[htbp]
	\centering
	\includegraphics[width=0.85\linewidth]{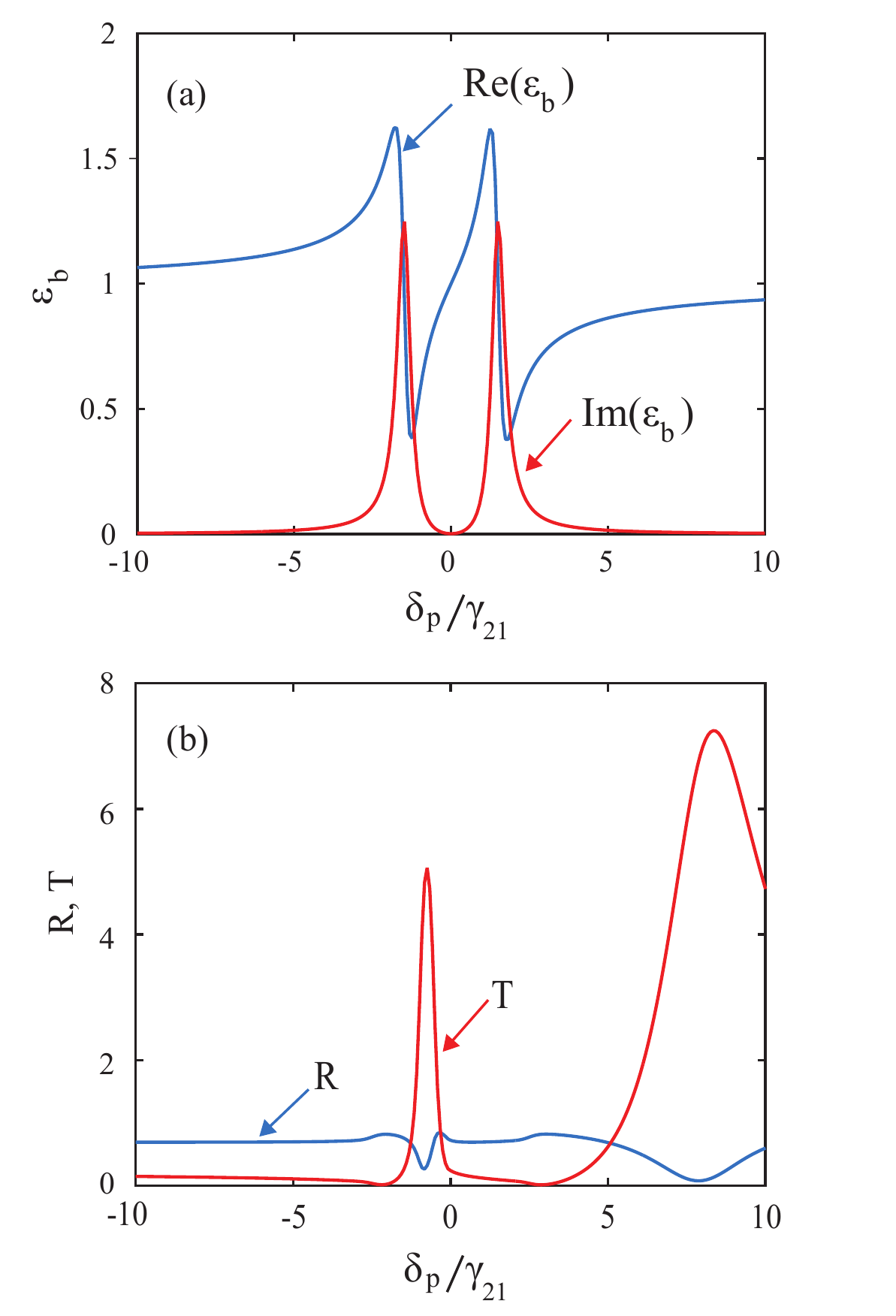}
	\caption{(a) The relative permittivity of the EIT medium (described by Eq. 3). $Re(\epsilon_{b})$ is represented by blue solid line while the $Im(\epsilon_{b})$ is represented by the red solid line, and (b) the transmittance (red line) and the reflectance (blue line) of the proposed three-layer structure. Here, we have taken the parameters: $\theta_{p}=80^{o}$, $\theta_{c}=0^{o}$, $q=25 \ nm$, $\gamma_{21}=61.54 \ MHz$,  $\Omega_{c}=3\gamma_{21}$ and $\lambda_{p}=589.1 \ nm$, $\omega_{31}= 1.8 \ GHz$ \cite{steck2000sodium}.} 
\end{figure}
\section{Coupler-free excitation of surface plasmons in the linear regime}
In the linear regime, the density matrix elements and the probe field could be expanded using the perturbation method \cite{van2015eit} as ${\rho_{ij}}= {\rho^{(0)}_{ij}}+{\rho^{(1)}_{ij}}$ and $\Omega_{p}(\vec{r},t)=\epsilon {\Omega^{(1)}_{p}}(\vec{r},t)$, respectively where $\epsilon<<1$ is the perturbation parameter. Moreover, we consider ${\Omega^{(1)}_{p}}(\vec{r},t)=Fe^{i\phi}$, and ${\rho^{(1)}_{ij}}(\vec{r},t)={\rho^{(1)}_{ij}}(\vec{r}) Fe^{i \phi}$, where $F$ is a constant describing the pulse envelope of the SPs and $\phi=K(\omega)z-\omega t$,  with $\omega$ as the frequency perturbation of the SP and the linear dispersion of the atomic medium is given by
\begin{equation} \label{eq:5}
K(\omega)=\frac{\omega}{n_{eff} c}+ \frac{\kappa (\omega+\delta_{c}+i\gamma_{31})}{ {|\Omega_{c}|}^{2}-(\omega+\delta_{p}+i\gamma_{21})(\omega+\delta_{c}+i\gamma_{31})}.
\end{equation}
By assuming a weak field limit of the incident probe field, we have ${\rho_{11}}^{(0)} \approx 1$, ${\rho^{(0)}_{22}} \approx 0$  and ${\rho^{(0)}_{33}} \approx 0$. Using these conditions, we can obtain the expression for ${\rho_{21}}$ to the first order as follows:
\begin{equation}\label{eq:6}
{\rho^{(1)}_{21}}(\vec{r})=\frac{{-i\Omega^{(1)}_{p}}\big[\gamma_{31}-i(\delta_{p}-\delta_{c}) \big]}
{2\Big[\big(\gamma-i\delta_{p}\big)\big[\gamma_{31}-i(\delta_{p}-\delta_{c})\big]+ {\big(\Omega_{c}/2 \big)}^{2} \Big]}. 
\end{equation} 
Then, the linear susceptibility of the atomic medium is given by 
\begin{equation}\label{eq:7}
{\chi^{(1)}_{p}}=\frac{-iN{|\vec{\mu}_{21}|}^{2}}{\hbar \epsilon_{0}} {\rho^{(1)}_{21}}. 
\end{equation}    

In the proposed scheme, the excitation of surface plasmon polaritons can be explained with the help of the transmittance $(T)$ and the reflectance $(R)$ of the three-layer structure. Following the method described in ref \cite{du2015coupler}, the transmittance and the total reflection is calculated using the expressions, $T={|t_{tmb}|}^{2}$ and $R={|r_{tmb}|}^{2}$, respectively, where $t_{tmb}$ and $r_{tmb}$ are the three-layer transmission and the reflection coefficients of the three-layer structure. The normal wave vectors is given by  ${k^{2}_{jz}}={k^{2}_{0}} \epsilon_{k}-{k^{2}_{x}}$ where $j = t, m, d$ while the parallel wave vector is given by $k_{x}=k_{0} n_{t} \sin \theta_{p}$, where $k_{0}=\omega / c$ is the vacuum wave number and $\theta_{p}$ is the incident angle of the probe field. The relative permittivity and the excitation of the SP resonance is depicted in Fig. 2. Figure 2(a) represents the real and imaginary part of the relative permittivity while Fig. 2(b) represents the transmittance and the reflectance of the proposed structure. Here, we consider the three-level atomic medium as the sodium D2 line $(3^{2} S_{1/2} \rightarrow 3^{2}P_{3/2})$, with the energy levels $\ket{1}=\ket{3S_{1/2}, F=1}$, $\ket{2}=\ket{3P_{3/2}, F=0}$ and $\ket{3}=\ket{3S_{1/2}, F=2}$, with a typical density of the range, $N= 10^{16}$ to $ 10^{18} \ m^{-3}$ \cite{steck2000sodium}. For EIT to occur, we assume that probe beam is much weaker than the pump beam. Here, we have taken the parameters \cite{steck2000sodium}: $\gamma_{21}=61.54 \ MHz$,  $\Omega_{c}=3 \gamma_{21}$, $\lambda_{p}=589.1 \ nm$, $\omega_{31}= 1.8 \ GHz$. Additionally, the middle layer is considred to be a layer of Silver metal having a dielectric permittivity, $\epsilon_{m}= -13.3+0.883i$ \cite{palik1998handbook} which is approximated to be constant within the EIT window. Then, for a particular value of the angle of incidence of the probe and the coupling field, the atomic medium exhibits the EIT phenomenon. It is evident from the Fig. 2(a), that the condition $Re(\epsilon_{b})<1$ and $Im(\epsilon_{b})<<1$ is achieved when $\delta_{p} \approx 0$. In this case, the relative permittivity, $\epsilon_{b}<1$ and as a result the wave-number of the SPs, $\beta=k_{0} \sqrt{\frac{\epsilon_{m} \epsilon_{b}}{ \epsilon_{m}+ \epsilon_{b}}}$ can be less than the wave-number of the incident light, i.e., $\beta < k_{0}$. Hence, the surface plasmon resonance (SPR) condition  which is given by the expression $k_{0} n_{t} \sin\theta_{p}= \beta$ can be satisfied even for $n_{t}=1$ for a proper incidence angle $\theta_p$ \cite{du2015coupler}. At this position, it is observed that the transmittance $T$ is greatly enhanced and the total reflection $R$ drops to a small value such that $T>>R$ (see Fig. 2(b)), which is the signature of the resonant excitation of SPs. 

Furthermore, we can investigate the group velocity of the excited SPs ($\tilde{V_{g}})$ for proper set of parameters of the incident laser fields. For the parameters: $\delta_p=0$, $ \delta_c=3 MHz$ and $\Omega_c=3 \gamma_{21}$, the value of group velocity $\tilde{V_{g}} \approx 0.97 c$, i.e., the excited SP waves travel with a subluminal group veolcity. This group velocity can be further tuned for proper set of parameters of the incident laser fields. It is noteworthy to mention that a reduced group velocity signifies an enhanced lifetime of the excited SPs in the proposed structure. Hence, a coupler-free excitation of subluminal SPs at the interface of the metal and the EIT medium is possible in the proposed structure. In the proposed coupler-free planar plasmonic waveguide struture, it is natural to examine the nonlinear behavior of the excited SP resonances. In the nonlinear regime, the SPs can exhibit interesting solitonic behavior which can be of great significance in overcoming the propagation loss inherent in the case of SPs. Therefore, in the subsequent sections, we further examine the generation and evolution of the surface polaritonic solitons and the surface polaritonic breathers in the proposed structure.

\section{Excitation of surface polaritonic solitons}
The dynamics of the coupler-free excited surface plasmon polaritons in the non-linear regime, can be investigated with the help of the standard multiple scales method \cite{huang2005dynamics}, by introducing the asymptotic expansions of the density matrix equations, $\rho_{ij}-{\rho^{(0)}_{ij}} = \sum_{l} \epsilon^{l}{\rho^{(l)}_{ij}}$ and the Rabi frequency of the probe field $\Omega_{p}=\sum_{l}\epsilon^{l}{\Omega^{l}_{p}}$ where $x_{l}=\epsilon^{l} x\ (l=0,1,2)$ and $t_{l}=\epsilon^{l}t \ (l=0,1)$ are the multiscale variables. By substituting these expansions in the Maxwell Bloch equations a linear and inhomogeneous set of equations for ${\rho^{(l)}_{ij}}$ and ${\Omega^{(l)}_{p}}$  is obtained, which can be solved for different orders. For the second order, the solvability condition leads to the propagation of the probe pulse, which can be expressed as
\begin{equation} \label{eq:8}
\Bigg(\frac{\partial}{\partial x_{1}}+ \frac{1}{ V_{g}}\frac{\partial}{\partial t_{1}}\Bigg)F=0
\end{equation}
where $V_{g}$ is the group velocity and F is the envelope function of the probe field which is yet to be determined. We can obtain the group velocity by using the relation, $V_{g}= 1/ K_{1}$ where $K_{1}= \partial K(\omega) / \partial (\omega)$ is the first-order dispersion. Further, the solvability condition in the third order is expressed as  		
\begin{equation} \label{eq:9}
i\frac{\partial}{\partial x_{2}}- \frac{1}{ 2} K_{2}\frac{\partial^{2}F}{\partial {t_{1}}^{2}}-W{|F|}^{2}F e^{-2\alpha x_{2}}=0
\end{equation}	
where $\alpha=\epsilon^{2} Im (K(\omega))$  represents the absorption of the atomic medium, $K_{2}=\partial^{2} K(\omega ) / \partial(\omega)^{2}$  represents the group velocity dispersion (GVD), and the Kerr nonlinearity $W \propto {\chi^{(3)}_{p}}$, where ${\chi^{(3)}_{p}}$ is the third order susceptibility given by 
\begin{equation}\label{eq:10}
{\chi^{(3)}_{p}}= \frac{iN{|\vec{\mu}_{21}|}^{4}}{ \hbar \epsilon_{0}} {\rho^{(3)}_{21}}.
\end{equation}
Here, we can obtain the expression for ${\rho_{21}}$ to the third order as follows
\begin{equation}\label{eq:11}
{\rho^{(3)}_{21}}=  \frac{i{\Omega^{(1)}_{p}}}{2D}\Bigg[\frac{{\big({\Omega^{(1)}_{p}}}\big)^2 }{2\gamma+\gamma_{21}} \Bigg( \frac{1}{D} +\frac{1}{D^*}\Bigg) + \frac{2\gamma_{31} }{2\gamma+\gamma_{21}} \Bigg]
\end{equation}
where $D= \big( \gamma-i\delta_{p}\big)-\frac{\Omega_{c}/2}{\big[\gamma_{31}-i\big(\delta_{p}-\delta_{c} \big)  \big]}$. By choosing the parameters: $\gamma_{21}=61.54 \ MHz$, $ \delta_{c}= 3 \ MHz$,  $\Omega_{c}=3 \gamma_{21}$, $\lambda_{p}=589.1 \ nm$ and $\omega_{31}= 1.8 \ GHz$, we obtain the value of susceptibility, which is described by Eq. 10, as ${\chi^{(3)}_{p}} = (5.55 + 0.0832i) \times 10^{-7} \ m^{2} V^{-2}$. Then, we get the Kerr coefficient as $
n_{2}= \chi^{(3)}_{p} /{2\sqrt{1+\chi^{(1)}_{p}}} \approx 2.7 \times 10^{-7} \ m^{2} V^{-2}$. This nonlinearity competes with the dispersion in the system giving rise to solitonic behavior of the SPs in the system. The nonlinear Schrodinger equation (NLSE), describing the dynamics of the SPs in the nonlinear regime, can be obtained by combining all the equations in all the orders, as follows: 	
\begin{equation} \label{eq:12}
i \Bigg(\frac{\partial}{\partial x}+ \alpha \Bigg) U - \frac{1}{ 2} K_{2}\frac{\partial^{2}U}{\partial T}-W{|U|}^{2}U=0,
\end{equation}	
where, $T=t-x⁄\tilde{V_{g}}$ , $U=\epsilon F e^{-\alpha x}$ and $\alpha=Im (K(\omega))$. At the EIT position, there is null absorption (i.e.,$\alpha \approx 0$) and the imaginary part of the coefficients is much smaller than its real parts. Here, by neglecting the imaginary parts of the coefficients, Eq. 12 reduces to the normalized form:
\begin{equation} \label{eq:13}
i \frac{\partial u}{\partial \zeta}+ \frac{1}{ 2} \frac{\partial^{2}u}{\partial \tau}+\delta{|u|}^{2}u=0,
\end{equation}
where $x=-L_{D} \zeta$, $T=T_{0} \tau$ and $u=U / U_{0}$. The dispersion length of the medium is given by $L_{D} = {T_{0}}^{2}/ \tilde{K_{2}}$ and the nonlinear length of the medium is given by $L_{non}=1/|{U_{0}}^{2}|\tilde{W}$. Here, $T_{0}$ is the pulse duration, $\delta $ denotes the Kerr nonlinearity and $ U_{0} = \frac{1} {T_{0}} \sqrt{{\tilde{K}}_{2}/ \tilde{W}}$ is the typical Rabi frequency of the probe field. For $\delta = +1$, Eq. 13 can assume the fundamental bright soliton solutions, which is given by $u_{B}=\sech(\tau)e^{i\zeta}$. The bright polaritonic soliton solution in the original variables is given by the expression: ${\Omega_{p}}^{B}=U \exp(i \tilde{K_{0}x})$, which can be expressed as:
\begin{equation}\label{eq:15}
	{\Omega^{B}_{p}}= \frac{1}{T_{0}} {\Bigg( \frac{{\tilde{K}_{2}}}{\tilde{W}} \Bigg)}^{2}  \sech \Bigg[\frac{1}{T_{0}} \Big(t-\frac{x}{\tilde{V_{g}}} \Big) \Bigg]  e^{i\big[\tilde{K_{0}}- \frac{1}{2L_{D}}   \big]x},                                          
	\end{equation}
	where the $\tilde{K_{2}}$ and $\tilde{W}$ denotes the real part of the variables $K_{2}$ and $W$. Then, the corresponding electric fields for the bright soliton can be expressed as:
	\begin{equation}\label{eq:16}
	{E^{B}_{p}}= \frac{\hbar}{{ |\vec{\mu}_{21}}|T_{0}} {\Bigg( \frac{{\tilde{K}_{2}}}{\tilde{W}} \Bigg)}^{2}  \sech \Bigg[\frac{1}{T_{0}} \Big(t-\frac{x}{\tilde{V_{g}}} \Big) \Bigg]  e^{i\big[\big(\tilde{K_{0}}+k_{p}- \frac{1}{2L_{D}} \big)x -\omega_{p}t \big]},                                          
	\end{equation}
	\begin{figure}[htbp]
		\centering
		\includegraphics[width=0.95\linewidth]{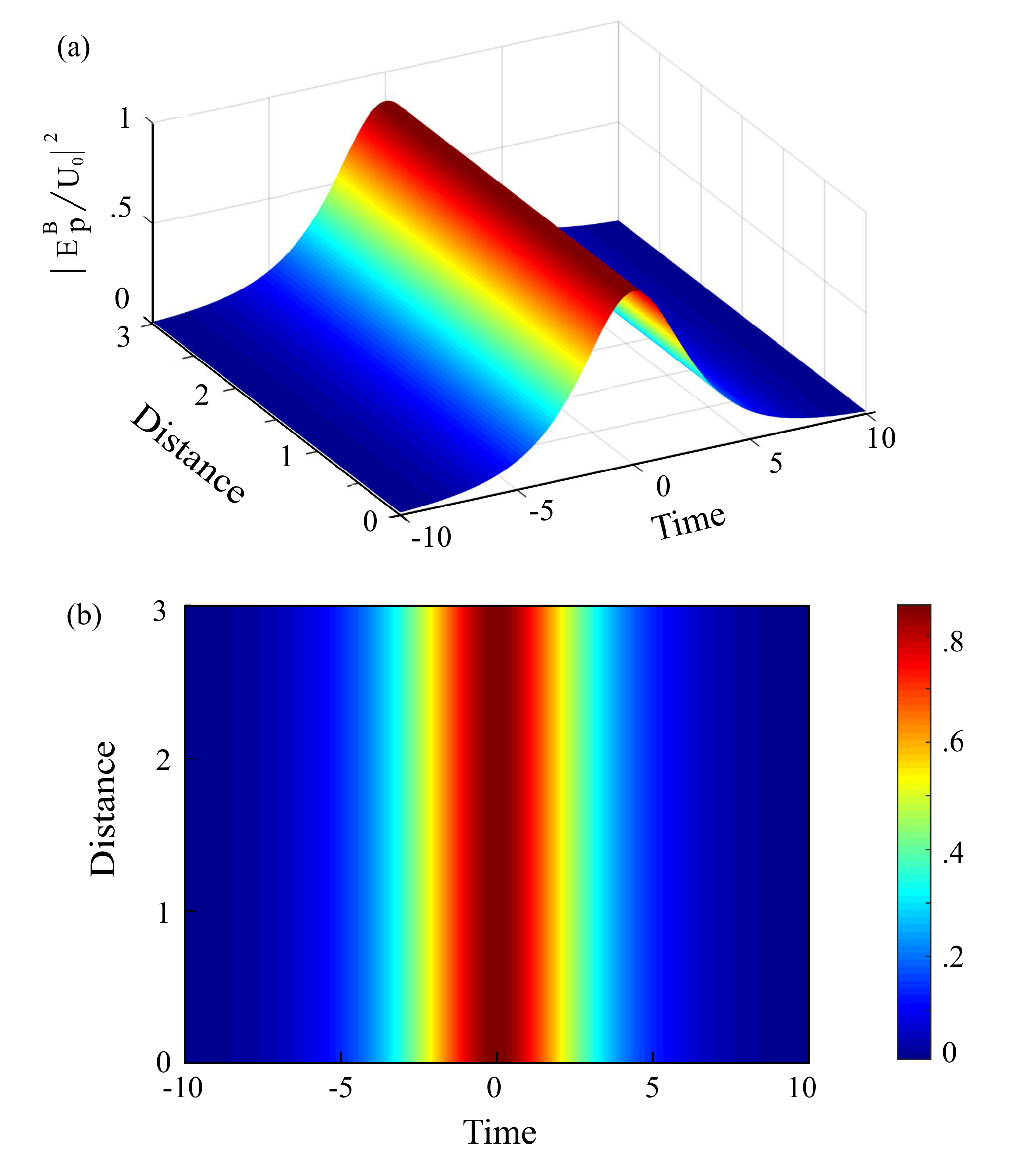}
		\caption{The spatiotemporal dynamics of the bright polaritonic solitons within the EIT transparency window, in the proposed structure as a function of distance and time. (a)The evolution of the probe field intensity ${|{E}^{B}_{p}/U_{0}|}^{2}$ as a function of  $x/L_{D}$ and $T/T_{0}$. (b) The 2D contour plot of the bright polaritonic solitons corresponding to (a) for the following parameters: $\delta_{c}= 3 \ MHz$, $T_{0}=10 \ ps$, $\Omega_c= 3 \gamma_{21} $, $\omega_{p}=3.19 \times 10^{15}\ Hz$.} 
	\end{figure}

In order to investigate the evolution and the propagation for the bright polaritonic solitons, we assume the initial condition as ${{E^{B}_{p}}(0,t)}/{U_{0}} =\sech \big({T}/{T_{0}} \big)exp \big(i{T}/{T_{0}}\big)$. It is noteworthy to mention that the proposed scheme can be observed in an experiment by choosing realistic parameter. By taking $\delta_{c}= 3 \ MHz$, $T_{0}=10 \ ps$, $\Omega_{c}=3 \gamma_{21}$, one can assume a stable propagation of the bright polaritonic soliton. For these specific values, we get $W= (1.47 + 0.069i) \times 10^{-16}\ m^{-1} s^{2}$. The spatiotemporal dynamics of the bright polaritonic solitons within the EIT transparency window, in the proposed structure as a function of distance and time is described in Fig. 3. Figure 3(a), illustrates the evolution of the probe field intensity ${|{E}^{B}_{p}/U_{0}|}^{2}$ as a function of $x/L_{D}$ and $T/T_{0}$ while Fig. 3(b) represents the corresponding 2D contour plot of the bright polaritonic solitons. It is evident from the figure that the bright polaritonic solitons can propagate through the waveguide for a long distance without encountering much distortion. The bright polaritonic soliton retains its shape and amplitude as it propagates through the system. This stable propagation is achieved as a result of the balanced dispersion with the sufficient self-focusing provided by the inherent Kerr nonlinearity within the EIT transparency window of the system. 
\begin{figure}[htbp]
	\includegraphics[width=0.95\linewidth]{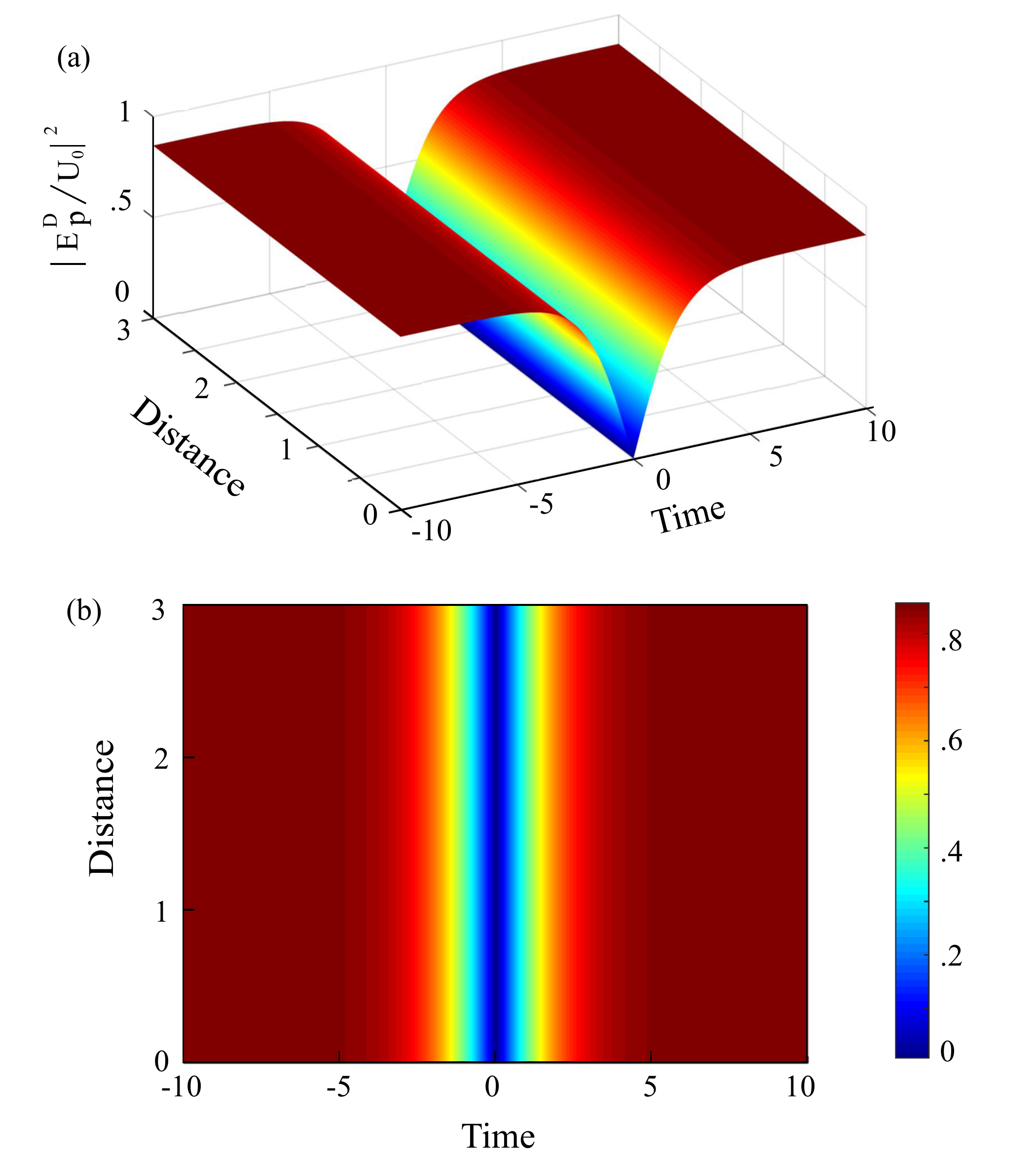}
	\caption{Excitation and propagation of the dark surface polaritonic solitons in the planar plasmonic waveguide structure. (a) The evolution of the probe field intensity ${|{E^{D}_{p}}/U_{0}|}^{2}$ as a function of  $x/L_{D}$ and $T/T_{0}$ along with (b) the corresponding contour map.}
\end{figure}

Similar investigation can be carried out for the case of dark polaritonic solitons as well. For $\delta = -1$, we obtain the fundamental dark soliton solutions which is given by $u_{D}=\tanh(\tau)e^{i\zeta}$. By returning to the original variables, the dark polaritonic soliton solution is given by the relation  ${\Omega^{D}_{p}}=U \exp(i\tilde{K_{0}}x)$, which can be further expressed as follows:
\begin{equation}\label{eq:17}
	{\Omega^{D}_{p}}= \frac{1}{T_{0}} {\Bigg( \frac{{\tilde{K}_{2}}}{\tilde{W}} \Bigg)}^{2}  \tanh \Bigg[\frac{1}{T_{0}} \Big(t-\frac{x}{\tilde{V_{g}}} \Big) \Bigg]  e^{i\big[\tilde{K_{0}}- \frac{1}{2L_{D}} \big]x },                                          
	\end{equation}
	For the dark polaritonic solitons, the corresponding electric fields is given by the expression:
	\begin{equation}\label{eq:18}
	{E^{D}_{p}}= \frac{\hbar}{{|\mu_{21}}|T_{0}} {\Bigg( \frac{{\tilde{K}_{2}}}{\tilde{W}} \Bigg)}^{2}  \tanh \Bigg[\frac{1}{T_{0}} \Big(t-\frac{x}{\tilde{V_{g}}} \Big) \Bigg]  e^{i\big[\big(\tilde{K_{0}}+k_{p}- \frac{1}{2L_{D}} \big)x -\omega_{p}t \big]},                                          
	\end{equation} 

The propagation dynamics of the dark polaritonic soliton is numerically investigated by taking the initial condition is taken as $
{{E^{D}_{p}}(0,t)}/{U_{0}} =\tanh \big({T}/{T_{0}} \big)exp \big(i{T}/{T_{0}}\big)$. Figure 4 illustrates the spatiotemporal evolution of the dark surface polaritonic solitons in the planar plasmonic waveguide structure. The evolution of the probe field intensity ${|{E}^{D}_{p}/U_{0}|}^{2}$ as a function of  $x/L_{D}$ and $T/T_{0}$ is shown in Fig. 4(a) along with the corresponding contour map in Fig 4(b). Similar to the case of the bright polaritonic solitons, the initial pulse retains its shape and amplitude for a long distance in the case of the dark polaritonic solitons as well, resulting in the stable propagation of the pulse in the proposed system. Such undistorted propagation of these polaritonic solitons can find immense applications in communication in the optical regime. 

\begin{figure*}[t]
	\centering
	\includegraphics[width=1\linewidth]{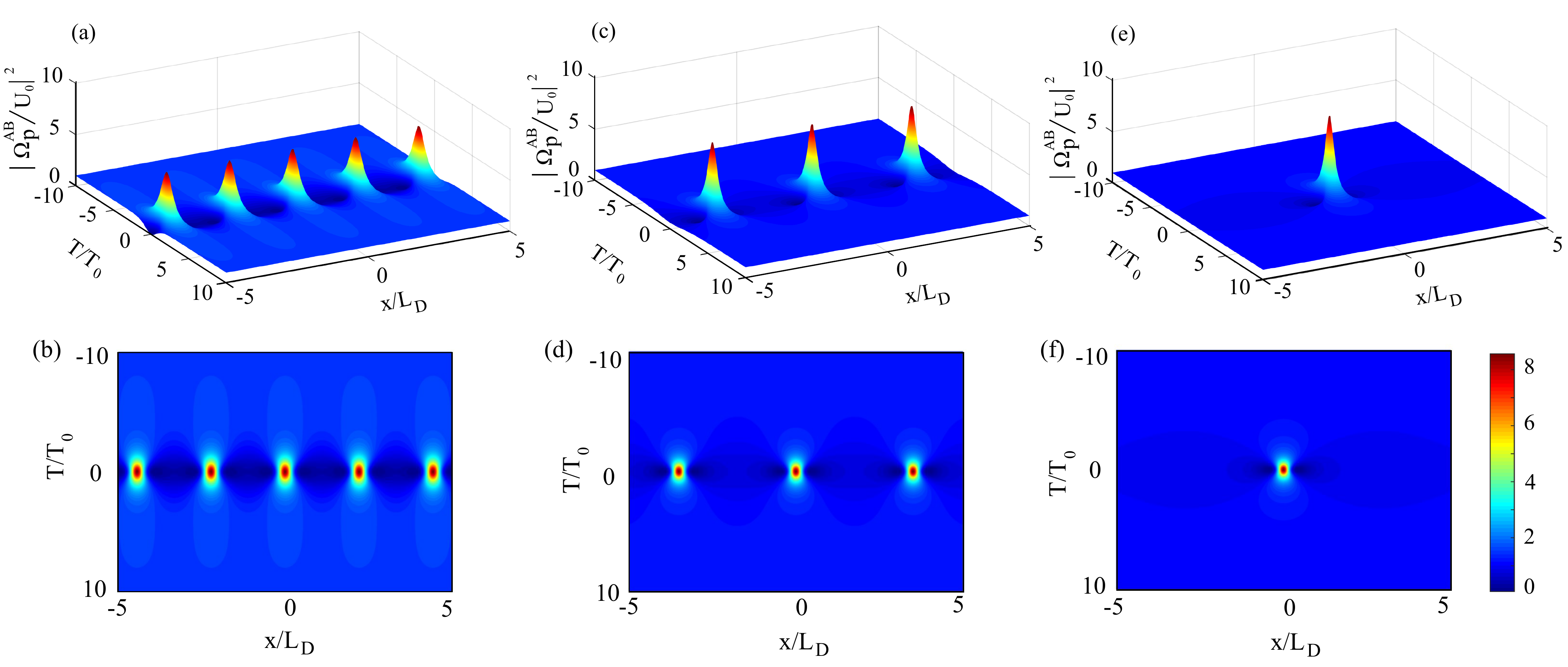}
	\caption{The dynamics of the surface polaritonic Akhmediev breathers in the planar plasmonic waveguide structure. (a) The evolution of the probe field intensity ${|{\Omega}^{AB}_{p}/U_{0}|}^{2}$ as a function of  $x/L_{D}$ and $T/T_{0}$ along with (b) the corresponding contour map. Here, we have used the parameters $a = 0.25$,  $\Omega = 1.41$, and $b = 1$.(c) The evolution of the surface polaritonic Akhmediev breathers along with its (d) corresponding contour map for the parameters $a = 0.4$,  $\Omega = 0.89$, and $b = 0.8$. (e) The Peregrine soliton along with its (f) corresponding contour map.} 
\end{figure*}
\section{ Excitation of surface polaritonic Akhmedeiv breathers}
The standard NLSE, described by Eq. 13, can assume other exact solutions which are periodic in $\zeta$ and $\tau$ . The Akhmediev breather is one such solution, and it is given by the expression
\begin{equation} \label{eq:19}
u=1+\frac{2[1-2a] \cosh(b\zeta)+ib\sinh(b\zeta)}{\sqrt{2a}\cos(\Omega\tau)-\cosh(b\zeta)}.
\end{equation}
	Then, we have 	
	\begin{equation} \label{eq:20}
	{\Omega^{AB}_{p}}=U_{0} \Bigg[1+\frac{2[1-2a] \cosh(b\zeta)+ib\sinh(b\zeta)}{\sqrt{2a}\cos(\Omega\tau)-\cosh(b\zeta)} \Bigg] e^{i\big[\tilde{K_{0}}x+\frac{x}{2L_{D}} \big]}.
	\end{equation}
	Returning to the original variables we get\\
\begin{equation}
\begin{split}
&{\Omega^{AB}_{p}}(x,T)= U_{0}\Bigg[1+\\
&\frac{[1-4a] \cosh(bx/L_{D})+\sqrt{2a}\cos(\Omega T/T_{0}) ib\sinh(bx/L_{D})}{\sqrt{2a}\cos(\Omega T/T_{0})-\cosh(bx/L_{D})} \Bigg]\\  
&e^{i\big[\tilde{K_{0}}x
	+\frac{x}{L_{D}} \big]},
\end{split}
\end{equation}
\\	
where, $a$ is the modulation parameter, $\Omega= 2 \sqrt{1-2a}$ is the spatial modulation frequency and $b= \sqrt{8a(1-2a)}$ is the parametric gain coefficient. $T= \pi / \sqrt{1-2a}$ is the time period of the periodic pulses. The spatiotemporal evolution of the surface polaritonic Akhmediev breathers in the planar plasmonic waveguide structure for different values of $a$, is depicted in Fig. 5. In order to have a physical understanding of the evolution of the surface polaritonic Akhmediev Breather, we can assume the initial SPs to be a plane wave (corresponding to $a=0$) which evolves into a significant pulse shape due to the parametric gain coefficient. As a result the SP’s amplitude gets significantly amplified by a growth factor of $ b$. With further modulation of the instability, the amplified SP’s evolves into a train of periodic pulses along the time axis with a period of $T$. Hence, the first-order surface polaritonic Akhmediev breathers can be potentially excited and propagated in this waveguide. 

The dynamics of the surface polaritonic Akhmediev breather in the proposed structure depends strongly on the spatial modulation frequency. Here, for $0 < a < 0.5$, it is evident from Fig. 5 that the spatial separation between the adjacent peak intensities increases with the increase of $a$. By taking realistic values of the modulation parameter as $a=0.25$ (see Figs. 5(a), (b)), we can obtain a train of pulses with a time-period $T = 1.4 \pi$, and $\Omega = 1.41$, $b = 1$. Figures 5 (c) and 5(d), represents the breather propagation for the modulation parameter, $a= 0.4$ with $T = 2.23\pi$, $\Omega = 0.89$ and $b = 0.8$. It is observed that the spatial width of the individual pulse decreases with increasing value of the modulation parameter while the temporal width of the pulses increases with an increase of the modulation parameter. Finally, for a limiting case of $a \rightarrow 0.5$, there is a significant spatial and temporal localization of the pulse leading to an increased localization of peak intensity. This spatiotemporally localized pulse is the so-called Peregrine soliton which is depicted in Figs. 5(e) and 5(f). Hence, a stable propagation of the surface polaritonic Akhmediev Breather is possible in the structure with judicious choice of parameters. The effective localization of these surface polaritonic breathers can lead to the generation of extremely short pulses in such coupler-free planar plasmonic structure. 

\section{Conclusion}
In conclusion, we investigate the excitation of surface polaritonic solitons and breathers in a coupler-free planar plasmonic waveguide structure comprising of a transparent layer, a metal layer and a three-level lambda-type atomic medium. In the linear regime, it is observed that the coupler-free excitation of SP resonances is possible in the proposed structure. Further, a giant Kerr nonlinearity is achieved in the system as a result of electromagnetically induced transparency in the bottom atomic layer, which can be controlled through proper modulation of the parameters of the driving laser fields. The self-phase modulation caused by the Kerr nonlinearity balances the group velocity dispersion in the system hence providing the necessary condition for the excitation of surface polaritonic solitons within the narrow transparency window of the electromagnetically induced transparency. It is observed that the system yields laterally self-trapped bright and dark surface polaritonic solitons which is tightly guided at the interface of the metal and the EIT medium. Finally, we have shown that a stable propagation and an effective localization of the surface polaritonic Akhmediev breathers can be achieved with proper choice of parameters in the proposed planar plasmonic waveguide structure. This study can be significant in the development of highly compact nano-optical and photonic devices for applications in the optical regime.

\end{document}